\newcount\style \style=2
\ifodd\style
 \documentstyle[aas2pp4]{article}
  \else
   \documentstyle[12pt,aasms4]{article}
\fi %\input psfig

\begin{document}
\setlength{\baselineskip}{12pt}

\newcommand\bb[1] {   \mbox{\boldmath{$#1$}}  }

\newcommand\del{\bb{\nabla}}
\newcommand\bcdot{\bb{\cdot}}
\newcommand\btimes{\bb{\times}}
\newcommand\BV{Brunt-V\"ais\"al\"a\ }

    \def\dd{\partial}
    \def\tilde{\widetilde}
    \def\etal{et al.}
    \def\eg{e.g. }
    \def\etc{{\it etc.}}
    \def\ie{i.e.}
    \def\beq{ \begin{equation} }
    \def\eeq{ \end{equation} }
    \def\spose#1{\hbox to 0pt{#1\hss}} % from Scott Tremaine
    \def\ltsim{\mathrel{\spose{\lower.5ex\hbox{$\mathchar"218$}}
	 \raise.4ex\hbox{$\mathchar"13C$}}}

\def\tilde{\widetilde}

\long\def\Ignore#1{\relax}

\title{Stability, Instability, and ``Backwards'' Transport\\
in Stratified Fluids.}

\author{Steven A.~Balbus}
\affil{Virginia Institute of Theoretical Astronomy, Department of Astronomy,}
\affil{University of Virginia, Charlottesville, VA 22903-0818}
\affil{sb@virginia.edu}

\vskip 2 truein
%\centerline{\today}
%\centerline{To appear in the {\sl Astrophysical Journal, {\bf 521,}} 20 August 1999.}
\vskip 2 truein

\begin{abstract}

The stratification of entropy and the stratification of angular
momentum are closely analogous.  The analogy has been developed for a
number of different problems in the fluid literature, but its
consequences for the behavior of turbulent accretion disks are less
appreciated.  Of particular interest is the behavior of disks in which
angular momentum transport is controlled by convection, and heat
transport by dynamical turbulence.  In both instances we argue that the
transport must proceed ``backwards'' relative to the sense one would
expect from a simple enhanced diffusion approach.  Reversed angular
momentum transport has already been seen in numerical simulations;
contra-gradient thermal diffusion should be amenable to numerical
verification as well.  These arguments also bear on the observed
nonlinear local stability of isolated Keplerian disks: locally generated
turbulence in such a disk would require simultaneous inward and outward
angular momentum transport, which is of course impossible.

We also describe a diffusive instability that is the entropy analogue
to the magnetorotational instability.  It affects thermally stratified
layers when Coulomb conduction and a weak magnetic field are present.
The plasma must be sufficiently dilute that heat is channeled only
along field lines.  The criterion for convective instability goes from
one of upwardly decreasing entropy to one of upwardly decreasing {\it
temperature.}  The instability remains formally viable if radiative
heat transport is also present, but the equilibrium is much more
unstable if Coulomb transport is dominant.  In that case, the maximum
growth rate is of order the inverse sound crossing time, independent of
the thermal conductivity.  The indifference of the growth rate to the
conduction coefficient, its simple dynamical scaling, and the
replacement in the stability criterion of a conserved quantity
(entropy) gradient by a free energy (temperature) gradient are
properties similar to those exhibited by the magnetorotational
instability.


\end{abstract}

\section{Introduction}

Although one normally associates the turbulent transport of heat with
thermal convection, and the turbulent transport of (angular) momentum
with shear instabilities, they are sometimes called upon to interchange
roles.  In the study of accretion disks, for example, two related
problems arise: does convective turbulence act like an enhanced
viscosity, and does MHD turbulence act like an enhanced thermal
diffusivity?  In what sense is turbulence compelled by irreversible
thermodynamics to transport from hot to cold and from fast to slow?  In
the course of this paper, we will see that great care is needed when
trying to answer such questions, and that nothing can be taken for
granted.

We shall make frequent use of a very deep analogy between angular
momentum and entropy stratification.  The analogy is well-known in the
fluid community, but seems less appreciated by disk theorists.  An
extensive review and applications to specific two-dimensional flows may
be found in the article by Veronis (1970), who credits the
original insight for the analogy to Lord Rayleigh, and there are
briefer discussions in the texts by Greenspan (1968) and Tritton
(1988).  

Our purpose here to show that the angular momentum--entropy
correspondence is a powerful conceptual tool for understanding aspects
of turbulent transport and flow stability.  Even in circumstances in
which the analogy is not a mathematical identity, it affords physical
insights.  For example, the magnetorotational instability is a
destabilizing process in which a magnetic field taps into the angular
velocity gradient, a source of free energy, in the presence of an
otherwise stabilizing angular momentum gradient.  Might there be a
related analogue instability which taps into a thermal gradient, also a
free energy source, in the presence of an otherwise stabilizing entropy
gradient?  The answer, we shall see, turns out to be yes.

Our understanding of the large scale properties of accretion disk
turbulence also is sharpened by seeking analogous dynamical and thermal
behavior.  Three-dimensional numerical simulations of convective
instability have consistently yielded inward angular momentum transport
(Stone \& Balbus 1996, Cabot 1996), opposite to the sense one would
expect if the turbulence behaved as an enhanced viscosity.  We shall
argue that vertical turbulent heat transport in an accretion disk must
also behave in the same contrary fashion: the dynamical agitation of a
convectively stable layer will drive heat from the cooler upper layers
to the hotter midplane layers.

This paper is organized as follows.  In \S 2, we present a simple
physical picture for inward angular momentum transport in convectively
unstable disks, and suggest a reason for why pressure gradients are
ineffective for causing outward transport.  Section 3 presents the
analogous story for downward heat transport in a convectively stable
vertical layer.  (In this case, of course, pressure gradients are
irrelevant.) Section 4 presents the temperature gradient analogue of
the magnetorotational instability, discussing both its power, its
limitations, and describing its physical basis.  Finally, \S 5
summarizes our findings.

\section {Angular Momentum Stratification}
\subsection {A Preliminary Comment}

The stability of Keplerian disks and the inward turbulent transport of
angular momentum are different manifestations of the same underlying
fluid behavior:  epicyclic mixing of disturbed fluid elements.  The
analogous behavior in a thermally stratified layer would correspond to
\BV oscillations and inward turbulent heat transport, as we discuss in
\S 3.

Consider an isolated gas disk, and erect a standard cylindrical
coordinate system $(R, \phi, z)$ with its origin at the disk center.
Turbulence is assumed to be present.  Let $\bb{u}$ be the residual
velocity of a fluid element with the circular motion $R\Omega$
subtracted out.  We shall refer to $\bb{u}$ as a ``fluctuation'', but it
may have a nonvanishing mean value.  In a thin Keplerian disk, these
mean values will tend to be much smaller than the rms value of the u
velocity magnitude (Balbus \& Hawley 1998).

The production of entropy by turbulence occurs if the stress
tensor $T_{R\phi}\equiv \langle \rho u_R u_\phi \rangle$ is positive,
i.e., if the angular momentum flux is outward.  (The angle brackets
denote a local spatial average of the fluctuations; $\rho$ is the mass density.)
This is a sensible
result, since the inward drift of mass, which releases (half) the
orbital binding energy to be radiated, is possible only if the material
sheds its angular momentum.  If mass goes inward, angular
momentum needs to go out.  Formally, the turbulent dissipation rate is given by
$-T_{R\phi} d\Omega/d\ln R$ (e.g. Shakura \& Sunyaev 1973,
Balbus \& Hawley 1998).  Requiring this quantity to be
positive implies $T_{R\phi} > 0$ whenever $\Omega$ decreases outward.  
%\beq\label{flucen}
%{\dd\ \over\dd t}  \left\langle {\cal E}\right\rangle + \del\bcdot
%\left\langle \bb{u} {\cal E} + \bb{u} P +\bb {F_{rad}} \right\rangle = -
%{d\Omega\over d\ln R}\left\langle\rho u_R u_\phi \right\rangle,
%\eeq
%where
%${\cal E}$ is the energy density 
%$$
%{\cal E} = {1\over2}\rho u^2 + {3\over2} P +\rho\Phi,
%$$
%$\bb{F_{rad}}$ is the radiative flux, and $\bb{u}$ is the velocity in excess of
%the local Keplerian value.  If we integrate over the volume of the disk, we
%obtain
%\beq\label{posT}
%\int \bb{F_{rad}} \bb{\cdot} \bb{dA} = -  \int 
%{d\Omega\over d\ln R}\left\langle\rho u_R u_\phi \right\rangle dV > 0,
%\eeq
%the radiated energy is positive, and therefore so must be the (shear weighted)
%volume integral of $T_{R\phi}$.  
The difficulty comes in reconciling this requirement
with the turbulent transport properties inherent to
Rayleigh-stable (hereafter R-stable) disks.

\subsection{Transport Phenomenology}

The issue involved may be grasped intuitively by a diagrammatic argument.
Fig. [1] shows a schematic plot of $\Omega(R)$ for a Rayleigh-unstable
(R-unstable) disk.
The points labeled $A$ and $B$ are fluid elements at two different
radii, say $R_A$ and $R_B$, $R_A< R_B$.  The dotted lines are the two
curves of constant angular momentum that pass through $A$ and $B$.  The
process of turbulent angular momentum transport involves mixing of the
two elements at some intermediate radius $R_I$, and two new elements
emerging with a different allocation of angular momentum.  The points
$A'$ and $B'$ represent the elements at the time just before their
interaction.  Let us assume that it is the nonlinear interaction of
fluid element mixing which causes a loss (or gain) of specific angular
momentum.  (We return to this critical point below.)  Then, the
movements from $A$ to $A'$ and from $B$ to $B'$ are nearly along curves
of constant angular momentum.  In this case, these curves are less
steep than the $\Omega(R)$ curve for the disk, since $\Omega(R)$
corresponds to an outwardly decreasing angular momentum profile.  We
denote the value of $\Omega$ at the diagram point $A$ as $\Omega_A$,
and similarly for the points $B, A', B'$.  Note that
$\Omega_A>\Omega_B$ and $\Omega_{A'} >\Omega_{B'}$; the last two
symbols represent the angular velocities of the displaced fluid
elements just before mixing.  We show below that the exchange is
irreversible in the sense that entropy is increased:  the interaction
conserves angular momentum but not mechanical energy, some fraction of
which must be lost as heat.  

In fig. [2], we illustrate the same problem for an R-stable disk.  In
this case, the constant angular momentum curves are steeper than the
actual $\Omega(R)$ disk curve, which we have suggestively labeled
$\Omega_{\rm Kep}$.  Notice that the topology of the curves has
changed completely.   Although $\Omega_A>\Omega_B$ as before, we now
have $\Omega_{A'} <\Omega_{B'}$.  This has important consequences.

Denote the post-interaction angular velocities with an asterisk:
$\Omega^*_{A'}$, $\Omega^*_{B'}$.  The actual post-interaction
{\em identities} of the fluid
elements are not important---the elements can mix to any degree.
But the $A$ and $B$ labels will always refer to the interior and exterior
element respectively.  
Angular momentum conservation gives
\beq\label{three}
\Omega_{A'} + \Omega_{B'}= \Omega^*_{A'} +  \Omega^*_{B'}.
\eeq
If a fraction $0 < \epsilon \ll 1$ of the kinetic energy is lost in the exchange, 
energy conservation gives
\beq\label{four}
(1-\epsilon)( \Omega_{A'}^2 + \Omega_{B'}^2) =
(\Omega^*_{A'})^2  +  (\Omega^*_{B'})^2.
\eeq
We calculate to first order in $\epsilon$, at which level any net mass
transport may be ignored.  Equations (\ref{three}) and (\ref{four})
yield to leading order
\beq\label{result}
\Omega^*_{A'} - \Omega_{A'} = {\epsilon E\over \Omega_{B'}- \Omega_{A'}}
\eeq
where 
\beq
E = {1\over2} ( \Omega_{A'}^2 + \Omega_{B'}^2).
\eeq
In an R-unstable disk, $\Omega_{B'}< \Omega_{A'}$, and since $\epsilon
> 0$ we arrive at a very sensible result, $\Omega^*_{A'} <
\Omega_{A'}$, angular momentum flows outward.  But for an R-stable
disk, $\Omega_{B'} > \Omega_{A'}$, and equation (\ref{result}) says
something more remarkable.  We must have $\Omega^*_{A'} > \Omega_{A'}$.  In
other words, angular momentum flows inward.

This is a counterintuitive but inescapable conclusion.  It is also a
result completely in accord with numerical findings (Kley, Papaloizou,
\& Lin 1993, Stone \& Balbus 1996, Cabot 1996).  In retrospect, the
ideal test of the inward transport prediction may be found in the
numerical simulations of convective disk turbulence.  In these
studies,  there is no {\em mechanical} forcing of the turbulence which
might give rise to some kind of external torque.  The turbulence is
thermally driven by heating from below.  Simulations not using
a large explicit viscosity show inward transport of angular momentum
(e.g., Stone \& Balbus 1996, fig.~[2]).  This can hardly be ascribed to
diffusive numerics.  Going to finer numerical resolution will not turn
the angular momentum flux around; if anything, it will make it
``worse'' (i.e., greater in magnitude in the inward direction), as any
small residual numerical diffusion is eliminated.  Inward transport is
not an accident of poor numerical resolution, it is an inevitable
large-scale epicyclic mixing phenomenon.

\subsection{The Inefficacy of Pressure Gradients}

The local azimuthal equation of motion for the $u$ velocity fluctuations 
is (Balbus \& Hawley 1998)
\beq\label{balbusaz}
{\dd\ \over\dd t}  \left\langle {\rho u_\phi^2\over2}\right\rangle
+ \del\bcdot\left\langle {1\over2}\rho u_\phi^2 \bb{u}\right\rangle =
-{\kappa^2\over2\Omega} \left\langle\rho u_R u_\phi\right\rangle
- \left\langle {u_\phi\over R}{\dd P\over\dd\phi} \right\rangle
-\eta_V \left\langle |\del u_\phi|^2 \right\rangle,
\eeq
where the angular brackets denote local averages, $P$ is the gas
pressure, $\kappa$ is the epicyclic frequency, and $\eta_V$ is the
dynamical viscosity.  (Note that despite the appearance of the angular
brackets, the only term dropped in the averaging procedure is one
associated with direct viscous transport.  Aside from neglecting
nonlocal curvature contributions, the inviscid terms are exact.)  To
maintain the $u_\phi$ fluctuations, we require a source term on the
right hand side of this equation.  The inward transport argument of the
previous section is evinced by the first source term on the right hand
side of equation (\ref{balbusaz}).  The second, pressure, term, could
in principle torque fluid elements so that specific angular momentum
conservation is violated.  Why doesn't this happen?

A key piece of information is the finding of Stone \& Balbus (1996)
that a uniformly rotating disk transports angular momentum inward at a
rate comparable to that found for a Keplerian disk.  This means that
inward transport is not simply a consequence of ``sheared out,'' nearly
axisymmetric flow structure.  Similar results are found when the shear
is zero.  There must be something intrinsic to the local retrograde
epicyclic response of a disk which precludes pressure gradients from
turning turbulent transport outwards.

There is.   It is manifested most readily in the simple behavior of
inertial waves in a uniformly rotating disk.  Inertial waves are the
fundamental linear response of an incompressible disk.   The Lagrangian
displacements of fluid elements in an axisymmetric wave is a retrograde
epicycle.  The inclusion of azimuthal pressure gradients does not alter
the basic character of the epicyclic response (nor does the presence of
vertical convection).  Out-of-phase components of motion are introduced
by pressure gradients, but they do nothing to change the systematic
tendency of a displaced element to move more slowly (rapidly) than
unperturbed neighboring disk orbits on an outward (inward) excursion.
The difficulty with pressure torques driving an outward turbulent
angular momentum is not that they are somehow too small, it is that
they are incoherent.  Indeed, in a barotropic fluid, pressure gradients
cannot alter the development of local epicyclic vorticity, and the
linear and nonlinear analysis differ only by a ``$\bb{v\cdot\nabla
v}$'' term.  This allows for mixing and mode coupling, but unless there
is complete disruption of the epicyclic response, it is not associated
with a change in the direction of the transport.  (None of this is a
concern for linear wave transport of course, whose directionality is
not determined by turbulent mixing.)

Epicyclic mixing need not occur, of course, for sufficiently large
imposed fluctuations.  Cabot \& Pollack (1992) have reported that
vigorous convection in a local simulation produced outward angular
momentum transport.  How much of this is due to the breakdown of
epicyclic mixing at sonic velocity scales, how much to secondary
density wave transport, and how much might be due to the artificially
low Reynolds number of the Navier-Stokes simulation (as the authors
themselves note) is at present unclear.  All numerical simulations of
relatively high Reynolds number, low Mach number convection reported in
the literature have thus far found inward transport.

A corollary of the transport arguments is that Keplerian disks cannot
be locally unstable to nonlinear perturbations.  For we have just
argued that such disturbances result in inward transport of angular
momentum in any R-stable disk.  But we have also noted in the beginning
of the section that turbulent dissipation must be accompanied by
outward angular momentum transport.  The transport cannot be both
inward and outward.  Therefore, turbulence cannot be sustained.  The
same conclusion must apply to convection in an isolated disk.  It is
impossible to self-consistently run a convective engine off of the
differential rotation of the disk, since we would once again be led to
the contradictory conclusion that the transport must be simultaneously
inward and outward.  But if we {\em externally} heat the disk at the
midplane, inward transport is an entirely self-consistent choice, one
that is consistently seen in numerical experiments.

\section{Turbulent Heat Transport in a Stratified Layer}

\subsection{Transport Phenomenology}

The ``dual'' problem in accretion disks of angular momentum transport
by convection is that of thermal transport by a dynamical shear
instability, presumably the magnetorotational instability.  The
argument of \S 2 applied to an entropy-stratified layer runs along very
similar lines.  In figs.~[3] and [4], we show two temperature profiles,
denoted $T_L(z)$.  Fig. [3] is convectively unstable (C-unstable), and
fig.~[4] is convectively stable (C-stable).  The dotted lines are
adiabats; they show $T(z)$ for an adiabatic equation of state for the
pressure profile of the layer.  In other words, they are the curves
followed by adiabatically displaced fluid elements in pressure
equilibrium with their surroundings.  If we interchange entropy for
angular momentum, and temperature for angular velocity, the topology of
the C-curves is exactly the same as the R-curves of \S\ 2.3.  In fact,
the graphs were drawn simply by replacing labels.

Consider the two fluid elements $A$ and $B$ shown in figs.~[3] and [4].
$A$ is initially hotter and lower than $B$.  The elements meet at the
same location $Z_I$ at the points $A'$ and $B'$, mixing and exchanging
heat, and separating.  In the C-unstable case shown in fig.~[3], at the
instant before mixing, $A'$ is hotter than $B'$.  But for the C-stable
case of fig.~[4], even though $A$ started out hotter than $B$, $A'$ is
{\em cooler\/} than $B'$: it has less entropy at the same pressure.
Whereas the C-unstable layer will transfer heat upwards, the C-stable
layer, if turbulent for some reason, will transfer heat {\em downwards.}

This has an immediate important consequence: there is no vertically
upward turbulent heat transport in a stably stratified accretion
disk, regardless of the level of turbulence.  The presence of magnetic
fields does not affect this conclusion, which is based solely on
thermodynamic arguments.   Non-turbulent transport (e.g. waves), or
highly compressive processes (e.g. shocks), need not be restricted in
this way.  But standard diffusive ``mixing-length'' transport theory
certainly is.

Let us consider the mixing more quantitatively.  The temperatures of
the elements will be denoted as $T_{A'}$, $T^*_{A'}$, etc., following
the labeling convention of \S\ 2.3.  
Thermal energy is conserved in the exchange,
\beq
T_{A'} + T_{B'} = T^*_{A'} + T^*_{B'}.
\eeq
Since the exchange increases entropy,
\beq
\ln \left( T^*_{A'}T^*_{B'}\over T_{A'} T_{B'}\right) = 
\epsilon > 0.
\eeq
If we assume $\epsilon \ll 1$, this states more simply
\beq
T^*_{A'}T^*_{B'} = T_{A'} T_{B'} (1+\epsilon).
\eeq
Solving 
equations (6) and (8)
to first order in $\epsilon$ for the temperature difference
$T^*_{A'} - T_{A'}$, we find
\beq
( T^*_{A'} - T_{A'}) \left( {1\over T_{A'}} - {1\over T_{B'}}\right)
= \epsilon > 0.
\eeq
If $T_{A'} > T_{B'}$, as in the C-unstable layer, then $T^*_{A'} <
T_{A'}$, and heat flows upward, as it should.   But in a C-stable layer, 
$T_{A'} < T_{B'}$; therefore $T^*_{A'} > T_{A'}$, and heat flows
downwards.  In either case, heat flows from the hot element to the
cooler, of course.  But adiabatic changes (expansion and compression)
put the hot element on top in a C-stable disk, even if the element is
initially cooler than the bottom dweller!  Heat flows down the
gradient of the ``potential temperature,'' (e.g. Tritton 1988), not of
the temperature itself.  

Notice how different turbulent heat diffusion is from the diffusion of a
passive contaminant, to which it is sometimes compared. Increasing the
rate of upward heat diffusion in a stably stratified gas layer is not
just a matter of agitation, any more than outward angular momentum
transport in a disk is a matter of heating the midplane.

\subsection {Statistical Argument}

There is a simple fluid
statistical argument to support the diagram analysis and the fluid
element mixing argument.  In the presence of thermal conduction, the
internal energy equation of an ideal gas may be written
\beq\label{entropy}
{d\ln T P^{{1\over\gamma} - 1}\over dt} = (1- {1\over\gamma})\, {1\over P}
\bb{\nabla \cdot}
\left( \chi \bb{\nabla}\ln T\right),
\eeq
where $\gamma$ is the specific heat ratio and
$\chi$ is the effective thermal conductivity.  We shall assume
that the temperature fluctuations in the gas $\theta \equiv
\delta\ln T$ are much larger
than the pressure fluctuations,
\beq
\theta \gg \delta \ln P,
\eeq
but not so large that the background entropy gradient becomes
ill-defined, i.e., $\theta  \ll 1$.  Take the fluctuating component of
eq.~(\ref{entropy}), multiply by $\theta$, and average.
Under these assumptions, the
temperature fluctuation equation becomes
\beq
{1\over2} {\dd\theta^2\over\dd t} + \langle \theta \bb{u}\rangle
\bb{\cdot \nabla}\ln TP^{{1\over\gamma} - 1}= -  \left( 1 - {1\over\gamma}\right)
{\chi\over P}
\langle |\bb{\nabla} \theta |^2\rangle,
\eeq
where the term on the right has been integrated by parts and the pure
divergence assumed to be small upon averaging.  To maintain
the temperature fluctuations at a finite level, one must have
\beq
\langle \theta \bb{u}\rangle \bb{\cdot \nabla}\ln TP^{{1\over\gamma} -1} < 0,
\eeq
i.e., in a C-stable gas layer, the $u_z\theta$ correlation must be
negative.  This is a statement of downward turbulent heat transport.

\section{Weak Field Instability of an Entropy Stratified Layer}

\subsection {Dispersion Relation}

One apparent asymmetry in the entropy $\leftrightarrow$ angular momentum
analogy is the existence of the magnetorotational instability.  A weak magnetic field
turns an R-stable disk into a morass of turbulence if the angular velocity (not the
angular momentum) decreases outward (Balbus \& Hawley 1991).
Is there anything analogous in an entropy stratified layer?

It is helpful to recall the spring model of the magnetorotational instability.  A
weak magnetic field acts as a spring-like couple between an inner and outer fluid
element.  Angular momentum flows from the inner to the outer, causing them to
separate.  But the increased separation raises the spring tension and the attendant
torque, resulting in a yet greater angular momentum transfer, so there is a runaway.

Return now to the thermal layer.  Substitute entropy for angular
momentum.  We need to have a mechanism that causes entropy to flow from
a lower to higher element.  The lower element loses entropy and sinks,
the upper element gains entropy and rises.  (We are careful to
distinguish entropy gains and losses from heating and cooling.)  The
difficulty is that we now require that the rate at which entropy flows
between the elements to increase as their separation increases.  What
entropy transport mechanism could possibly behave this way?

Consider a static, stratified {\em plasma\/} layer, with an upwardly decreasing
temperature profile, $dT/dz<0$.   All fluid variables depend only upon
$z$.  There is a finite Coulomb conductivity in the plasma ($\chi_C$),
but we also allow for the possibility for a non-Coulombic conductivity
as well, which for convenience we shall refer to as ``radiative,''
($\chi_R$).  There is a weak uniform magnetic field $B$ pointing in the
$x$ direction.  The magnetic field has no dynamical effect upon the
equilibrium, but we allow for the perturbations to exert Lorentz forces.
In this problem, the primary role of the magnetic field is to channel
heat along its lines of force:  cross field thermal conductivity is
unimportant.  The Larmor ion radius, which is about $(T/B)^{1/2}$ cm ($T$
in Kelvins, $B$ in Gauss), is assumed to be tiny.  Finally, in what
follows, we assume that the resistivity is small compared with the
thermal diffusivity, as would be the case for a hot, dilute plasma.

Let $\bb{\hat b}$ be a unit vector in the magnetic field direction.  The
Coulombic heat flux
due to the temperature gradient $\bb{\nabla}T$ is
\beq
\bb{Q_C} = -\chi_C \bb{\hat b}\> \bb{\hat b}\bb{\cdot\nabla}T,
\eeq
where $\chi_C$, as noted, is the Coulomb conductivity (e.g., Spitzer 1962). 
In the equilibrium state, $\bb{Q_C}$ vanishes.  To linear order, perturbations
($\delta$ notation) introduce a Coulombic heat flux
\beq\label{deltaq}
\delta \bb{Q_C} = - \chi_C \bb{\hat b}\> \delta \bb{\hat b}\bb{\cdot\nabla}T
- \chi_C \bb{\hat b} \bb{\hat b}\bb{\cdot\nabla}\delta T.
\eeq
The total heat flux is
\beq
\delta \bb{Q} = - \chi_C \bb{\hat b}\> \delta \bb{\hat b}\bb{\cdot\nabla}T
- \left(\chi_R\bb{\nabla} + \chi_C \bb{\hat b} \bb{\hat b}\bb{\cdot\nabla}
\right) \delta T.
\eeq
(Terms in $\delta \chi_{{ }_{(C,R)}}$ are higher order in a WKB analysis.)

We consider plane wave perturbations of the form $\exp(\sigma t +ikx)$,
the simplest possible plane wave form prone to the instability.  Only a
$z$ velocity perturbation is present ($\delta v_z$), which implies that
$\delta P$ must vanish.  Likewise, only a $z$ magnetic field
perturbation is present ($\delta B_z$).  The basic equations needed for
the nonvanishing $\delta$ quantities in our analysis are the vertical
components of the equation of motion and the induction equation, and
the entropy equation.  In linearized form, these equations read
\beq\label{lindyn}
\sigma \, \delta v_z = {\delta\rho\over\rho^2}{\dd P\over \dd z}
+{ikB\over4\pi\rho}\, \delta B_z,
\eeq
\beq\label{linind}
\sigma \, \delta B_z = ikB \, \delta v_z,
\eeq
\beq\label{linent}
- {\gamma}\sigma{\delta\rho\over\rho} + \delta v_z\, {\dd\ln
P\rho^{-\gamma}\over \dd z} = - (\gamma -1) {1\over P} \bb{\nabla\cdot} \delta\bb{Q}.
\eeq
We also have
\beq\label{deltab}
\delta \bb{\hat b} = {\delta B_z\over B} \bb{e_z},
\eeq
where $\bb{e_z}$ is a unit vector in the $z$ direction,
and
\beq\label{deltat}
{\delta T\over T} = - {\delta \rho\over \rho},
\eeq
since $\delta P = 0$.   Finally, let us denote
\beq
\chi'_C = {\gamma - 1 \over P} \, \chi_C, \qquad \chi' = {\gamma -
1\over P} \, ( \chi_C + \chi_R)
\eeq
The induction equation and dynamical equation together immediately
yield
\beq\label{lyndynbis}
\left( \sigma +{k^2v_A^2\over \sigma}\right) \, \delta v_z =
{\delta\rho\over\rho}\, {\dd P\over \dd z},
\eeq
where the Alfv\'en speed $v_A$ is as usual given by 
$v_A^2 = B^2/4\pi\rho$.  
Combining equations (\ref{linent}),
(\ref{linind}), (\ref{deltaq}), (\ref{deltab}), and (\ref{deltat})
leads to
\beq
- ({\gamma}\sigma + \chi' T  k^2 ){\delta\rho\over\rho} +
\delta v_z \left(
{\dd\ln P\rho^{-\gamma}\over \dd z} + {\chi'_C k^2\over \sigma}{\dd
T\over \dd z} \right) =0.
\eeq
Substituting for $\delta v_z$ from equation (\ref{lindyn}) and simplifying, 
we arrive at the dispersion relation,
\beq\label{instabc}
\sigma^3 + {\sigma^2\over\gamma} \chi' T k^2 + (N^2 +k^2 v_A^2) \sigma +
{k^2\over\gamma} T\left( \chi' k^2 v_A^2
-  {\chi'_C \over\rho} {\dd P\over \dd z}{\dd\ln  T\over \dd z}\right)
=0,
\eeq
where $N^2$ is the \BV frequency,
\beq
N^2 = - {1\over \gamma\rho} {\dd P \over \dd z} {\dd \ln P \rho^{-\gamma}\over
\dd z}. 
\eeq

\subsection {Weak Field Limit}

The instability is easiest to understand 
in the limit of a very weak field,
when its role is limited exclusively to channeling the heat flux.
If in equation ({\ref {instabc}), 
we drop all terms in $v_A^2$, the dispersion relation becomes
\beq\label {smallb}
\sigma^3 + {\sigma^2\over\gamma} \chi' T k^2 + N^2 \sigma -
{k^2\over\gamma} {\chi'_C \over\rho} {\dd P\over \dd z}{\dd\ln  T\over \dd z}
=0.  
\eeq
The effect of the magnetic field is hidden in the form adopted for the
magnetized conductivity, equation (14).   The   
small $\sigma$ solution of equation (\ref{smallb}) takes the form
\beq\label{smalltsig}
\sigma \rightarrow  - \chi'_C k^2 {\dd T\over \dd z}\left(\dd\ln P\rho^{-\gamma}
\over \dd z\right)^{-1}.
\eeq
This expression is real and positive when the entropy gradient and the
temperature gradient have opposite signs, in accord with the physical
description put forth in the beginning of this section.  Notice the
very important point that in a C-stable medium, the criterion for
stability changes from one of entropy increasing outward to one of
temperature increasing outward.  Once again, we see behavior analogous
to angular momentum stratification.  In an R-stable disk, the presence
of a magnetic field changes the stability criterion by replacing
angular momentum (conserved quantity) gradient with angular velocity
gradient (free energy source).  Here, in a C-stable layer we replace
entropy (conserved quantity) gradient with temperature gradient (free
energy source) to arrive at the correct stability criterion.

In fact, all of the omitted details of the initial description can now
be filled in.  The unexplained gap in our picture was how the rate of
entropy transfer increases with increasing separation of the fluid
elements.  As the elements separate, they take magnetic field lines
with them, aligning these heat conduits ever more parallel to the
background temperature gradient.  In this way, the effective conducting
component of the temperature gradient grows in proportion to the
separation, and so therefore does the rate of entropy transfer.  Note
that the lower element loses heat but gains in temperature, and that
the reverse is true for the upper element.  The point is that the lower
(upper) element is always cooler (warmer) that the surroundings it is
passing through, even as its temperature rises (falls).  This is an
example of a negative heat capacity effect, and it is exactly analogous
to an element losing angular momentum but gaining angular velocity in
an accretion disk.   In both instances, the behavior is the basis of a
runaway instability.

Using the expedient of solving for $\chi'_C T k^2$ in terms of $\sigma$,
it is straightforward to extract the maximum growth rate from the
dispersion relation (\ref{smallb}).  The dominant terms become the
second (quadratic) and the final (constant).  The result is
\beq\label{maxc}
\sigma^2_{max} = {\chi'_C\over\rho\chi'} \>
{\dd P\over\dd z}{\dd \ln T\over \dd z}.
\eeq
The only dependence on conduction is via the ratio $\chi'_C/\chi'$.
Hence, if only Coulomb conduction is present, the growth rate is
entirely independent of the thermal conductivity coefficient.
An simple illustrative numerical example with $\chi'=\chi'_C$ is
presented in fig.~[5], in which all three real branches of the
dispersion relation may be seen.  We have plotted $\sigma$ versus $\chi
' T k^2$, in units of the \BV frequency.  Motivated by symmetry
arguments, it is tempting to advance a conjecture similar to one
proposed for angular momentum stratified disks (Balbus \& Hawley 1992),
which has proven robust and useful: the growth rate of equation
(\ref{maxc}) is the largest possible growth rate any instability able
to tap into an outwardly decreasing temperature gradient in a C-stable
layer can achieve.  In practice, however, the role of radiative
conduction can be quite stabilizing, as we shall presently see.

\subsection {General Behavior of the Instability}

The quickest route to a general instability criterion for equation (\ref{instabc})
makes use of the {\it Routh-Hurwitz\/} criterion (e.g., Levinson \& Redheffer 1970).
Applied to our problem, the criterion states that all the roots of
the $\sigma$ dispersion polynomial have negative real parts if and only if
\beq
\left| \begin{array} {ccc}
 a_2 &   1  & 0\\
 a_0 &  a_1 & a_2\\
 0   &  0   & a_0
\end{array}   \right|  > 0,
 \eeq
where element $a_n$ in the determinant is the coefficient of $\sigma^n$
in equation (\ref{instabc}).
This leads to the {\it stability\/} requirement:
\beq\label{stability}
(\chi' k^2 v_A^2 - {\chi'_C \over\rho} {\dd P\over \dd z} {\dd\ln T\over \dd z})
(N^2 + { \chi'_C/\chi'\over\rho} {\dd P\over \dd z} {\dd\ln T\over \dd z}) >0.
\eeq
Now, it is easily shown that
\beq
N^2 + {1\over\rho}  {\dd P\over \dd z} {\dd\ln T\over \dd z} = 
(1 - {1\over\gamma}) {P\over\rho} \left( \dd\ln P\over \dd z\right)^2 >0.
\eeq
Assuming that $N^2 > 0$, so that the layer is convectively stable,
we must also then have
\beq
N^2 + {\chi'_C/\chi' \over\rho}  {\dd P\over \dd z} {\dd\ln T\over \dd z} >0
\eeq
since $\chi'_C/\chi'<1$.   Since $k$ may in principle be arbitrarily small,
the stability condition (\ref{stability}) reduces to
\beq
{\dd P\over \dd z} {\dd\ln T\over \dd z} < 0.
\eeq
This is in agreement with the maximum growth rate we obtained (eq.
[\ref{maxc}]) by ignoring the dynamical effects of magnetic fields.
Furthermore, we know that if the above criterion is violated,
nonpropagating local instabilities exist, since they emerge from a
small $\sigma$ analysis.  Thus we find quite generally that if the
temperature increases in the direction of the gravitational field, the
stratification is unstable.

A finite wavenumber disturbance is {\it unstable\/} if
\beq
k^2 v_A^2 - {\chi'_C \over\rho \chi'} {\dd P\over \dd z} {\dd\ln T\over \dd z}
< 0 \qquad {\rm (INSTABILITY)}
\eeq
which may be compared with the instability criterion of the magnetorotational
instability in its simplest form (Balbus \& Hawley 1998):
\beq
k^2 v_A^2 + {d\Omega^2\over d\ln R} < 0
\eeq
where the second term on the left is the radial gradient of the square
of the angular velocity.  The similarity is evident.  In a stratified layer,
magnetic tension competes with what amounts to a destabilizing buoyancy
force; in a disk it competes with a local destabilizing tidal force.  

An important practical limitation of the instability
is the ratio $\chi'_C/\chi'$, which can be very
small when radiative transport is operative:
\beq
{\chi'_C\over \chi'} \simeq 3 \times 10^{-5} \left( T\over 3\times
10^6\right)^{-1/2} \left( \kappa\rho\over \ln \Lambda\right).
\eeq
here, $\kappa$ is the opacity (not epicyclic frequency!), and
$\ln\Lambda$ is the Coulomb logarithm (Spitzer 1962).  Under
stellar-like conditions, if we restrict wavelengths to be smaller
than the characteristic pressure or temperature scale height (we do
not draw a distinction here), the magnetic field energy would have
to be a tiny fraction ($\sim 10^{-7}$) of the thermal energy to
trigger instability.  A more promising astrophysical venue for this
instability is the X-ray emitting plasma found in clusters of galaxies
(Markevitch et al. 1999), provided that attention is restricted to
cases of inwardly increasing temperature gradients.

Our purpose here, however, is not so much to find immediate utility as
it is to elucidate an interesting and surprising fluid instability---one
whose properties are a testament to the power and utility
of the angular momentum--entropy fluid analogy.

\section {Summary}

By drawing upon an analogy between the stratification of entropy and
angular momentum, we have been able to link peculiar aspects of
``backwards'' angular momentum transport in Keplerian disks with their
nonlinear stability, to make predictions of how nonconvective
turbulence will transport heat, and most surprisingly, to uncover a
new, potentially interesting, instability besetting
weakly magnetized thermal layers.

Transport in disks or thermal layers is never truly backwards, in the
sense of angular momentum flowing from slower rotation to more rapid
rotation, or of heat flowing from cold to hot.  The second law of
thermodynamics is explicitly satisfied in our picture.  The behavior
seems peculiar only if one views turbulence as some sort of large scale
kinetic theory with blobs standing in for molecules, completely
uncoupled from the constraints imposed by large scale flow structure.
The key idea underlying the analysis of the direction of heat transport
is that the entropy of an element is conserved as the fluid makes its
excursions; this is why the turbulent heat correlation couples to the
entropy gradient.  Obviously this behavior cannot be captured with a
kinetic theory, which can involve only gradients in molecular
velocities or their rms averages.

Transport appears to be backwards in an R-stable disk subject to
external thermal agitation because the higher angular momentum outer
fluid elements are rotating more rapidly than the lower angular
momentum fluid elements at the radius at which they mix.  Transport
appears to be backwards in an agitated C-stable thermal layer because
higher entropy outer fluid elements are hotter than the lower entropy
inner fluid elements at the point where they mix together.  This is the
content of fig.~[1--4].    But in both cases, transport flows in the
only direction compatible with an increase in entropy in the mixing
process.

The nonlinear stability of an isolated Keplerian disk is an immediate
consequence.  Outward transport angular momentum is essential to the
evolution of a disk, but inward transport is inevitable if fluid
elements exchange their angular momentum predominantly by mixing.  Such
disks cannot help but be stable in the absence of external heating or
driving.  The same reasoning excludes self-sustaining thermal
convection as a source of outward angular momentum transport.

Neither reversed angular momentum transport nor its thermal counterpart
seem likely to be very large; the $\alpha$ values of measured inward
convective transport tend to be of order $10^{-5}$ (Stone \& Balbus
1996).  It would be surprising if the reversed turbulent thermal
diffusivity due to MHD turbulence were competitive with upward
radiative diffusion as a heat transport mechanism: it is difficult and
inefficient to force heat downwards (as it is to force angular momentum
inwards).  While a numerical calculation is needed to be certain,
standard radiative diffusion accretion disk models are unlikely to be
strongly affected by the presence of MHD turbulence.

Perhaps the most unexpected finding of the entropy--angular momentum
analogy is the existence of a diffusive instability which operates in
magnetized thermal layers.  It relies upon heat flow restricted to
magnetic lines of force and assumes resistivity to be small, hence is
most appropriate for dilute plasmas.  The basis of the behavior of the
instability is very similar to that of the magnetorotational
instability:  in both cases transport is increased between neighboring
fluid elements as their separation is increased.  This drives the outer
element farther out and the inner element farther in.  In the case of
the thermal layer, heat transport causes a buoyant runaway; in the
magnetorotational instability, angular momentum transport causes a
dynamical runaway.  The increase in the entropy transport comes from
the magnetic field conduits aligning themselves more parallel with the
large scale temperature gradient as the displacement increases.
Disturbed fluid elements display negative heat capacity behavior: their
temperature and entropy fluctuations have opposite signs.  In the
angular momentum case, the increased transport is due to increasing
magnetic tension as the elements separate.   The fluid elements have
opposite angular momentum and angular velocity variations.

We have not discussed the nonlinear outcome of the instability, but
evidently it can be suppressed by allowing the temperature gradient to
follow magnetic lines of force.  If this is not possible, either
because of field line topology constraints or because field lines
transverse to the thermal gradient are somehow regenerated, then
turbulent convective transport may result when the temperature, not the
entropy, increases in the direction of gravity.

\section*{Acknowledgements}

I am indebted to an anonymous referee who drew my attention to 
papers from the fluid literature on the topic of the
rotation--stratification analogy, and for constructive advice.  I would
also like to thank C.~Terquem for her comments following a detailed
reading of an early draft of this paper, which led to a much improved
presentation, W.~Winters for his skillful preparation of the figures,
J.~Hawley, J.~Papaloizou, and C. Sarazin for useful conversations, and R.
Patterson for help with LaTex arcana.  This work was supported by NASA
grants NAG 5--3058 and NAG 5--7500.

\newpage
%\setcounter{totalnumber}{0}
%\renewcommand{\topfraction}{0.05}
%\suppressfloats
\begin{figure}
\epsscale{1.05}
\plotone{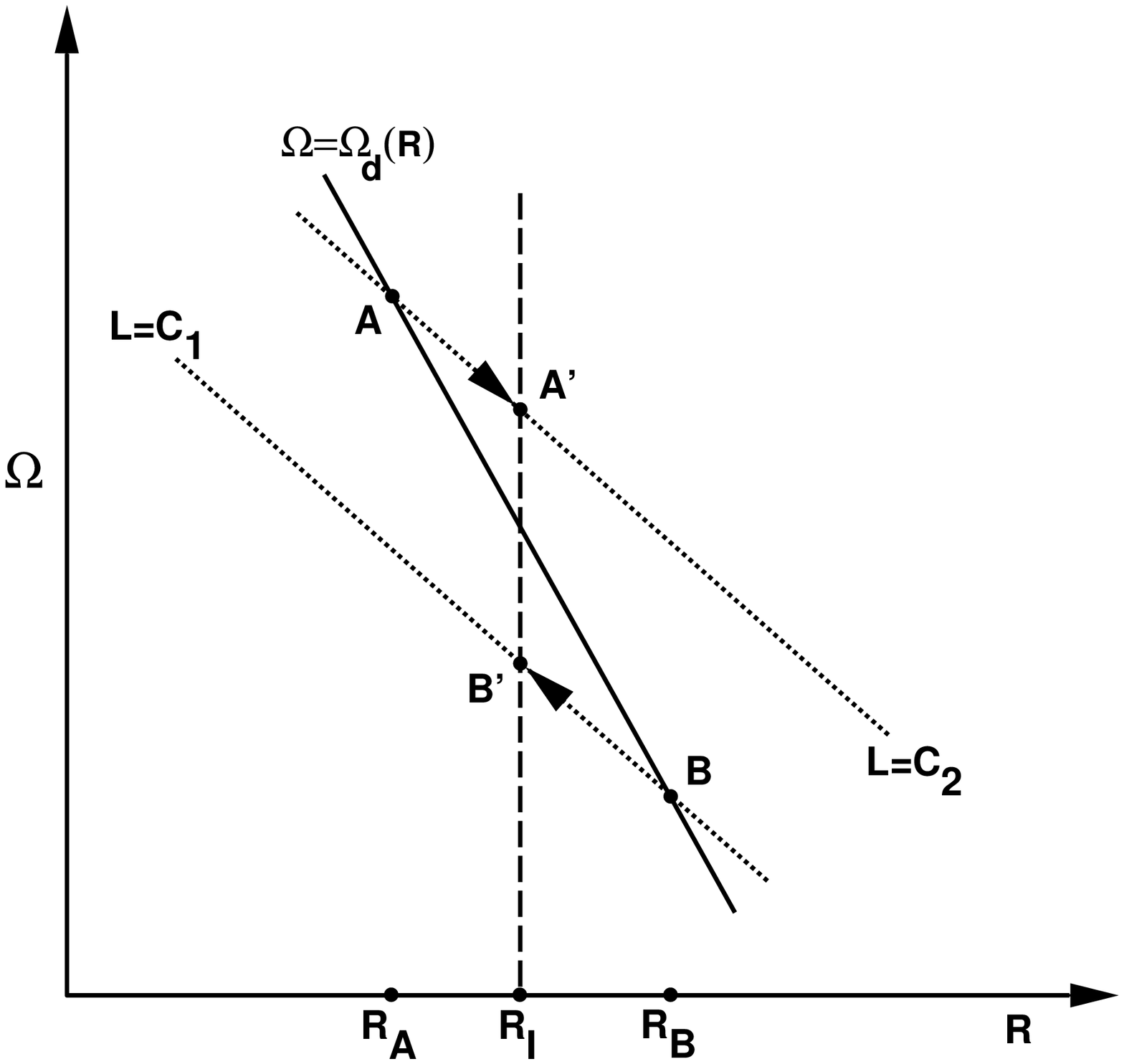}
\caption{
Representation of mixing in a Rayleigh-unstable disk.  When disturbed,
the fluid elements $A$ (inner) and $B$ (outer) tend to conserve their
angular momentum, moving along $L=$ constant paths $(C_1, C_2)$ to $A'$
and $B'$, until mixing at some intermediate location $R_I$.  Since $A'$
has geater angular momentum than $B'$ at the same location, it has
greater angular velocity, and angular momentum is transferred outward.
} 
\end{figure}

%\suppressfloats
\begin{figure}
\epsscale{1.05}
\plotone{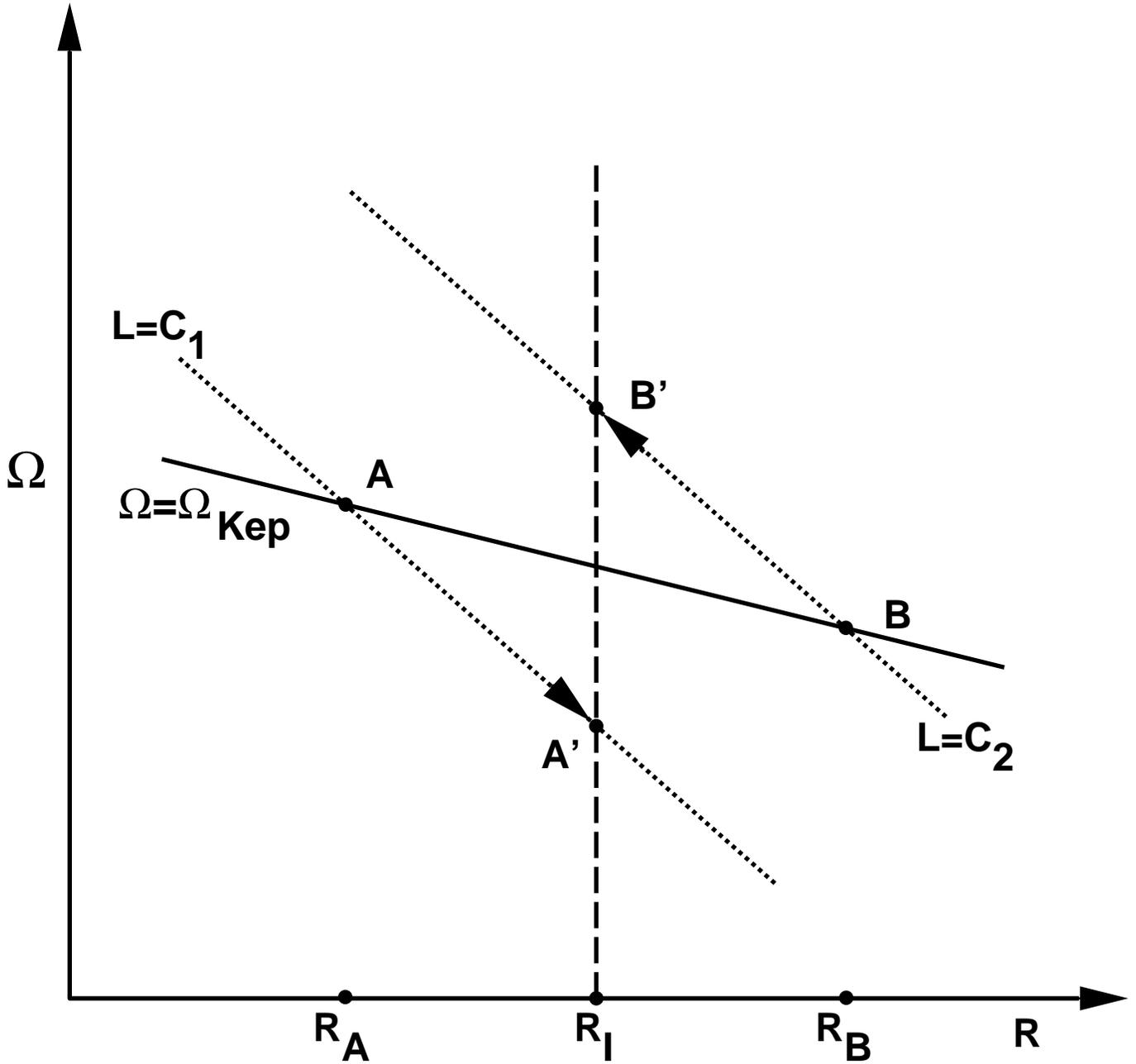}
\caption{
Representation of mixing in a Rayleigh-stable, Keplerian-like disk.  Since, in
contrast to fig.~[1], angular momentum increases outward in the disk,
$A'$ now has a less angular velocity than $B'$.  When the elements
mix, angular momentum is transferred {\em inward.}
}
\end{figure}

%\suppressfloats
\begin{figure}
\epsscale{1.05}
\plotone{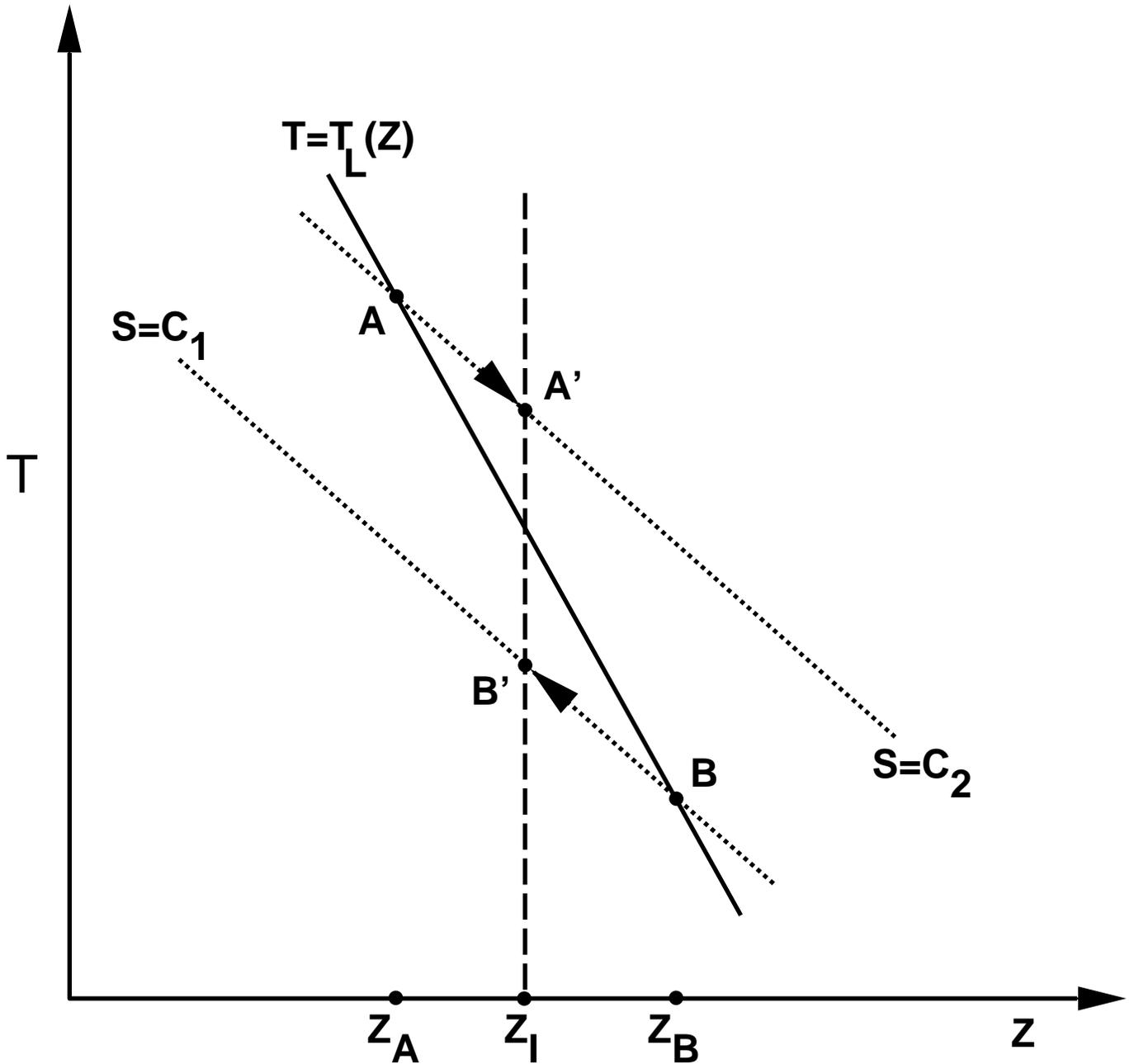}
\caption{
Representation of mixing in a convectively unstable layer.  When
disturbed, the fluid elements $A$ (lower) and $B$ (upper) tend to
conserve their entropy, moving along $S=$ constant paths $(C_1, C_2)$
to $A'$ and $B'$, until mixing at some intermediate location $Z_I$.
Since $A'$ has greater entropy than $B'$ at the same common external
pressure, it has a higher temperature, and heat is transferred
upward. 
} 
\end{figure}

%\suppressfloats
\begin{figure}
\epsscale{1.05}
\plotone{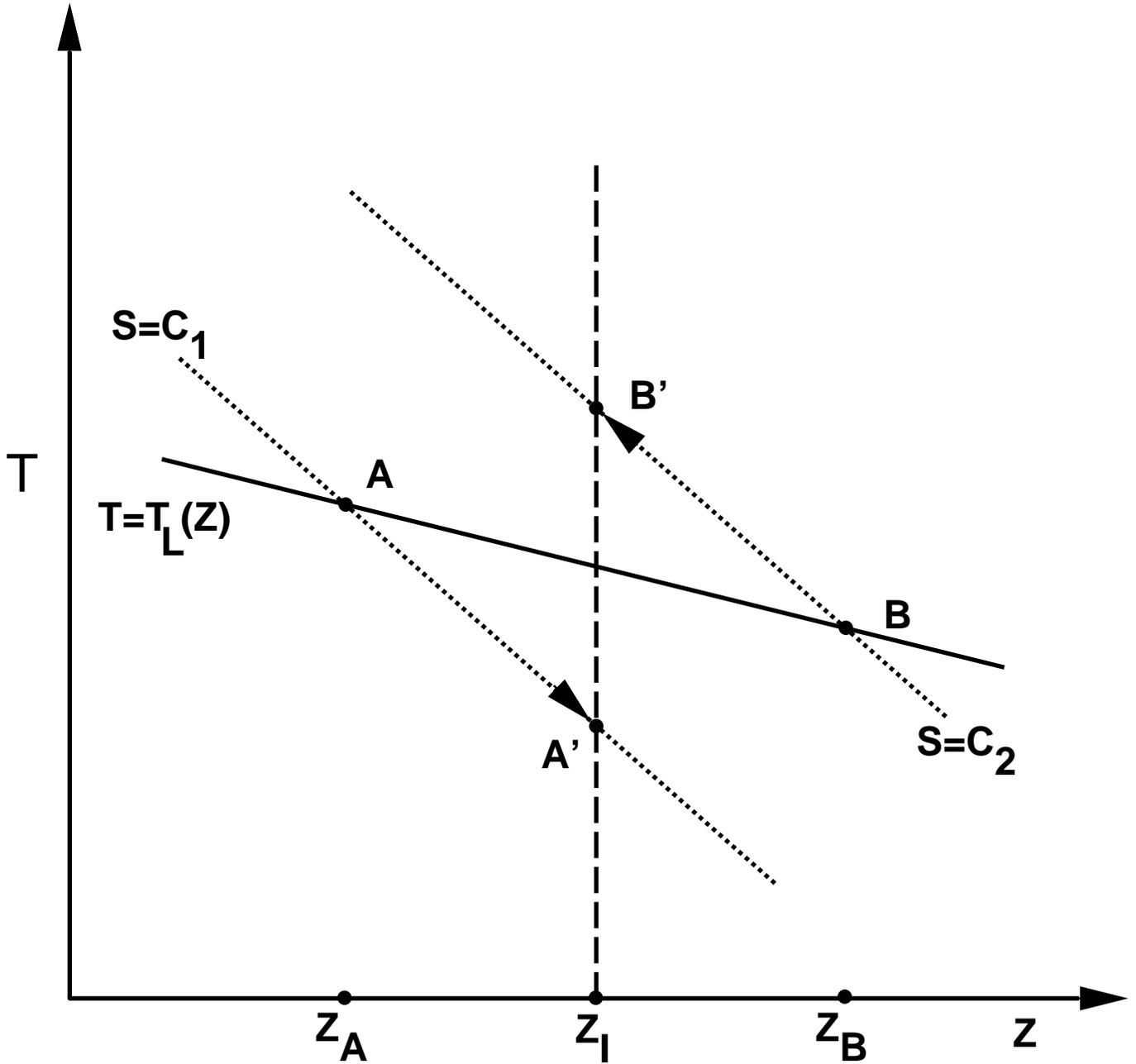}
\caption{
Representation of mixing in a convectively stable layer.  
Since, in contrast to fig.~[3],
entropy increases upward in the layer, $A'$ now has a smaller
temperature than $B'$.  When the elements mix, heat is
transferred {\em downward.}
}
\end{figure}

%\suppressfloats
\begin{figure}
%\epsscale {0.5}
%\plotone{back5.eps}
\plotfiddle{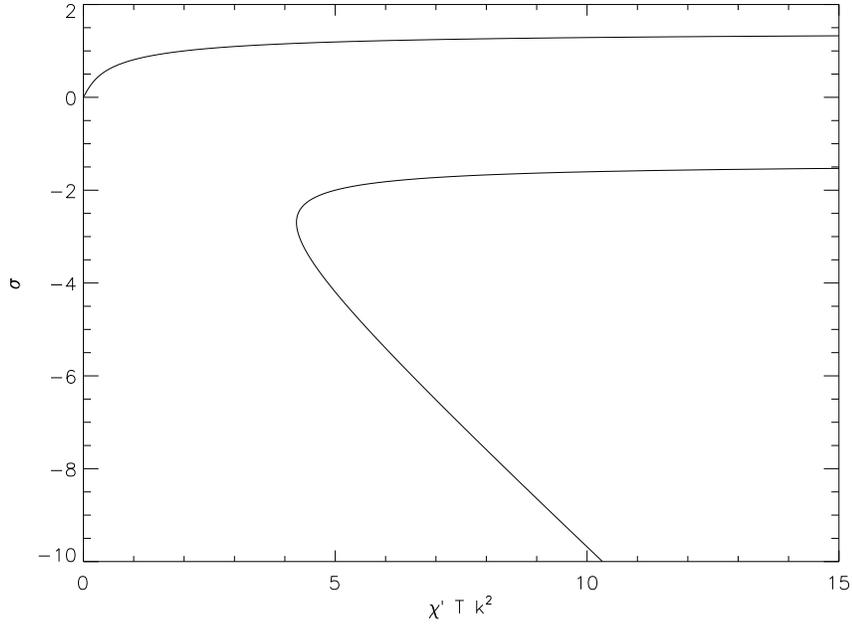} {3.0 truein} {90}{50}{50}{180}{0}
\caption{
The growth/damping rate for a stratified thermal layer in the presence
of a very weak magnetic field and finite thermal conductivity.  All
three branches are shown.  The magnetic field is assumed to restrict
heat flow parallel to lines of force, but to have no dynamical
consequences.  The units of the axes are in terms of the local \BV
frequency.  The example shown corresponds to vertical profile with
$d\ln T/d\ln P\rho^{-5/3} = -2$.
}
\end{figure}
%\suppressfloats

%\begin{figure}
%\plotfiddle{nu_disp.eps} {3.0 truein} {90}{50}{50}{180}{0}
%%\plotone{nu_disp.eps}
%\caption{The growth rate for a Keplerian accretion disk in the presence of a
%very weak magnetic field and finite ion viscosity.  The magnetic field is
%assumed to restrict momentum flow parallel to lines of force, but to have no
%dynamical consequences.  The units of the axes are in terms of the local
%angular velocity.  The maximum growth rate is $0.75 \Omega$, the Oort A-value
%of a Keplerian disk.}
%\end{figure}
%

\end{document}